\documentclass[twocolumn,english,twocolumn,showpacs,preprintnumbers,amsmath,amssymb,superscriptaddress]{revtex4}
\usepackage[T1]{fontenc}
\usepackage[latin1]{inputenc}
\usepackage{verbatim}
\usepackage{amsmath}
\usepackage{graphicx}
\usepackage{esint}

\makeatletter

\providecommand{\tabularnewline}{\\}

\@ifundefined{textcolor}{}
{%
 \definecolor{BLACK}{gray}{0}
 \definecolor{WHITE}{gray}{1}
 \definecolor{RED}{rgb}{1,0,0}
 \definecolor{GREEN}{rgb}{0,1,0}
 \definecolor{BLUE}{rgb}{0,0,1}
 \definecolor{CYAN}{cmyk}{1,0,0,0}
 \definecolor{MAGENTA}{cmyk}{0,1,0,0}
 \definecolor{YELLOW}{cmyk}{0,0,1,0}
}

\usepackage{babel}

\usepackage{babel}

\usepackage{babel}

\makeatother

\usepackage{babel}
\begin{document}

\title{Anisotropic emission of thermal dielectrons from Au+Au collisions
at $\sqrt{s_{NN}}=200$~GeV with EPOS3}

\author{Sheng-Xu Liu}

\affiliation{Institute of Particle Physics and Key laboratory of Quark \& Lepton
Physics (Ministry of Education), Central China Normal University,
Wuhan, China}

\author{Fu-Ming Liu}

\affiliation{Institute of Particle Physics and Key laboratory of Quark \& Lepton
Physics (Ministry of Education), Central China Normal University,
Wuhan, China}

\author{Klaus Werner}

\affiliation{Laboratoire SUBATECH, University of Nantes - IN2P3/CNRS - Ecole desMines,
Nantes, France}

\author{Meng Yue}

\affiliation{Institute of Particle Physics and Key laboratory of Quark \& Lepton
Physics (Ministry of Education), Central China Normal University,
Wuhan, China}

\date{\today}
\begin{abstract}
Dileptons, as an electromagnetic probe, are crucial to study the properties
of a Quark-Gluon Plasma (QGP) created in heavy ion collisions. We
calculated the invariant mass spectra and the anisotropic emission
of thermal dielectrons from Au+Au collisions at the Relativistic Heavy
Ion Collider (RHIC) energy $\sqrt{s_{NN}}=200$~GeV based on EPOS3.
This approach provides a realistic (3+1)-dimensional event-by-event
viscous hydrodynamic description of the expanding hot and dense matter
with a very particular initial condition, and a large set of hadron
data and direct photons (besides $v_{2}$ and $v_{3}$ !) can be successfully
reproduced. Thermal dilepton emission from both the QGP phase and
the hadronic gas are considered, with the emission rates based on
Lattice QCD and a vector meson model, respectively. We find that the
computed invariant mass spectra (thermal contribution + STAR cocktail)
can reproduce the measured ones from STAR at different centralities.
Different compared to other model predictions, the obtained elliptic
flow of thermal dileptons is larger than the STAR measurement referring
to all dileptons. We observe a clear centrality dependence of thermal
dilepton not only for elliptic flow $v_{2}$ but also for higher orders.
At a given centrality, $v_{n}$ of thermal dileptons decreases monotonically
with $n$ for $2\leq n\leq5$. 
\end{abstract}
\maketitle

\section{INTRODUCTION}

Dielectron spectra have been measured by the PHENIX Collaboration
\cite{PHENIX_PRC} and the STAR Collaboration \cite{STAR_PRL} in
Au+Au collisions at $\sqrt{s_{NN}}=200$~GeV at Relativistic Heavy
Ion Collider (RHIC). In fact, dileptons have been proposed as a kind
of clear probe in the study of the hot and dense matter named Quark-Gluon
Plasma (QGP), created in heavy ion collisions \cite{Shuryak}, due
to the fact that dileptons do not interact strongly and therefore
survive the whole evolution of a heavy ion collisions. This concerns
the initial hard scattering Drell-Yan pairs, the thermal emission
from the QGP and the Hadron Gas phase, and the decay of light and
heavy flavor hadrons. The dielectron spectra from both STAR and PHENIX
show the low mass enhancement beyond the cocktail. This phenomenon
was also observed in the CERES dielectron \cite{CERES_PLB,CERES_EPJ}
and NA60 dimuon data \cite{NA60_EPJ}. Medium modifications of the
$\rho$ meson spectral functions have been considered to describe
this enhancement, based on hadronic many body effective theories \cite{Rapp_1,Rapp_2,Rapp_3}
and using a relatively simple expanding fireball model. Another approach
used the forward scattering amplitude method \cite{Eletsky,Vujanovic}
and viscous hydrodynamics with smooth initial conditions as medium.
These two methods could get a similar result \cite{Rapp4}. The combination
of these two methods \cite{Xu1} could also describe the STAR dielectrons
data reasonably well, using a 2+1 dimensional ideal hydrodynamic model.
In addition, a microscopic transport model- Parton-Hadron String Dynamics
(PHSD) \cite{Linnyk1,Linnyk2,Cassing} has successfully described
the observed dimuon enhancement by NA60 and dielectron enhancement
by STAR.

Elliptic flow is the second Fourier coefficient of the azimuthal particle
distribution. The thermal elliptic flows of dileptons has been studied
in hydrodynamic simulations with smooth initial conditions \cite{Rupta,Mohanty}
and also with fluctuating initial conditions \cite{Xu2}. The fluctuations
seem to play an important role in elliptic flows of dileptons in heavy
ion collisions. However, the predicted elliptic flows of thermal dileptons
are small, so do their predicted elliptic flows of direct photons.
The observed large elliptic flow of direct photons seems a big puzzle
in our field. Using EPOS3102, a (3+1)D viscous hydrodynamic model
well constrained by hadron data, it seems that one can reproduce naturally
the measured elliptic and trianger flow of direct photons\cite{Fu2015}.
In this paper we will compute thermal dilepton production in an approach
which has been successfully tested concerning hadron and photon production.
This will serve as a benchmark to a systematic compilation of hadron,
dilepton and photon results.

In additional to the thermal sources, dileptons may be produced from
non-thermal sources such as primordial Drell-Yan annihilation and
electromagnetic final-state decays of long-lived hadrons. To account
for these, STAR cocktail results from Refs~\cite{STAR_PRL} have
been employed in this paper.

The paper is organized as following. In section 2, we briefly review
the calculation approach of thermal dilepton emission rates from quark-gluon
plasma (QGP) and hadronic matter. In section 3, the results will be
presented. Finally, a conclusion is made in section 4.

\section{CALCULATION APPROACH}

The calculation of dilepton production is obtained as an integration
over an emission rate $\frac{dR}{d^{4}q}=\frac{dN_{ll}}{d^{4}xd^{4}q}$,
based on the space-time evolution of a QGP and the hadronic stage
according to EPOS3\cite{epos3}. We use a QGP emission rate from extrapolating
lattice QCD results to finite momenta. The calculation of the hadronic
emission rate is done by employing the vector meson dominance model.
We consider explicitly the $\rho$ meson, for the rest we refer to
the so-called cocktail according to the STAR experiment (dileptons
from decays of measured resonances)~\cite{STAR2}. This procedure
to compute dileptions, discussed in more detail in the following,
will be referred to as EPOS3+DiL1.

\subsection{The EPOS3 dynamics}

As explained in \cite{epos3}, EPOS3 is an event generator based on
a 3+1D viscous hydrodynamical evolution starting from flux tube initial
conditions, which are generated in the Gribov-Regge multiple scattering
framework \cite{eposbas}. An individual scattering is referred to
as Pomeron, identified with a parton ladder, eventually showing up
as flux tubes (or strings). Each parton ladder is composed of a pQCD
hard process, plus initial and final state linear parton emission.
Nonlinear effects are considered by using saturation scales $Q_{s}$,
depending on the energy and the number of participants connected to
the Pomeron in question.

The final state partonic system (corresponding to a Pomeron) amounts
to (usually two) color flux tubes, being mainly longitudinal, with
transversely moving pieces carrying the $p_{t}$ of the partons from
hard scatterings \cite{eposbas,epos2}. One has two flux tubes, based
on the cylindrical topology of the Pomerons, but each quark-antiquark
pair in the parton ladder will cut a string into two, in this sense
one may have more than two flux tubes. In any case, these flux tubes
constitute eventually both bulk matter, also referred to as ``core''
\cite{eposcore} (which thermalizes, flows, and finally hadronizes)
and jets (also referred to as ``corona''), according to some criteria
based on the energy of the string segments and the local string density.


Concerning the core, we use a 3+1D viscous hydrodynamic approach,
employing a realistic equation of state, compatible with lQCD results.
We employ for all calculations in this paper a value of $\eta/S=0.08$.
Whenever a temperature of $T_{H}=168~$MeV is reached, we apply the
usual Cooper-Frye procedure to convert the fluid into particles. From
this point on, we apply a hadronic cascade \cite{urqmd}, based on
hadronic cross sections.

So we have two phases, a plasma phase with a hydrodynamical expansion,
and a hadronic phase, with individual hadron-hadron collisions. From
the fluid dynamical expansion of the plasma, we get the complete space-time
information, i.e. the collective velocity $\vec{v}(x)$ (which define
the local rest frame (LRF) at each point $x$) and the temperature
$T(x)$ of matter for a given space-time $x$, starting from some
initial proper time $\tau_{0}$. This is the basis of the dilepton
(rate) calculations discussed later. In the hadronic phase, we have
from EPOS a complete description of hadron trajectories. We may use
this to compute energy densities, flow velocities and net baryon density.
Assuming that the system may be approximated by a resonance gas in
equilibrium, we use the corresponding equation of state, c.f. Apendix
C of \cite{Werner:2010aa}, to get the temperature $T$ and baryon
chemical potential $\mu_{B}$ at any point in space time. This will
be used for the dilepton rate calculations. 

\subsection{Invariant mass spectrum}

The invariant mass spectrum of thermal dileptons is calculated as
an integration of the dilepton emission rates over the realistic space-time
evolution of the heavy-ion collision system and the momentum space
of the dileptons \cite{Rapp_1}, namely, 
\begin{equation}
\frac{dN_{ll}}{dM}(M)=\int d^{4}x\,\frac{Md^{3}q}{q_{0}}\,\frac{dN_{ll}}{d^{4}xd^{4}q}(T(x),q^{*}).\label{eq:intR}
\end{equation}
Here, $M$ and $q$ are the invariant mass and four momentum of the
dilepton pair, satisfying $M^{2}=q^{2}$. The four momentum $q^{*}$
of the dilepton pair as viewed from local rest frame is given as 
\begin{equation}
q_{0}^{*}=\gamma q_{0}-\gamma\vec{q}\cdot\vec{v}\label{eq:lorentztr1}
\end{equation}
and 
\begin{equation}
\vec{q^{*}}=\vec{q}+(\gamma-1)(\vec{q}\cdot\vec{v})\cdot\vec{v}/v^{2}-\gamma q^{0}\vec{v},\label{eq:lorentztr2}
\end{equation}
where $\vec{v}$ is the flow velocity, $v^{2}=\vec{v}\cdot\vec{v}$
and $\gamma=1/\sqrt{1-v^{2}/c^{2}}$. The flow velocity $\vec{v}(x)$
and the temperature $T(x)$ at each space-time point $x$ are provided
by the above-mentioned EPOS3 dynamics. The momentum integral ranges
from 0 to 5GeV/c (results do not depend on the upper bound as long
as it is much bigger than the temperature). The quantity $d^{4}x=\tau d\tau dxdyd\eta$
is the space-time element.

The space-time integral starts from the initial time $\tau_{0}$ of
the hydrodynamical evolution up to the end of hadron cascade. In the
hadronic phase, uRQMD has been employed after freeze-out in EPOS,
and macroscopic quantities such as energy-momentum and flow velocity
are reconstructed in order to calculate thermal emissions of dileptons
and photons, as mentioned earlier. The emission rate is the number
of dileptons emitted per space-time unit and per energy-momentum unit.
Thermal dileptons emitted from both the QGP phase and the hadron gas
phase are considered, as explained in the following.

\subsection{Dilepton Emission Rate in the Quark-Gluon Plasma}

For dileptons emitted from the QGP phase, asymptotic freedom implies
that the production rate in the intermediate invariant mass region
($1.<M<3.0GeV/c^{2}$) at high temperatures and densities can be described
by perturbation theory as in Ref.\cite{Rapp_1}. Recent progress using
thermal lattice QCD (lQCD) to calculate dilepton rates nonperturbatively
at vanishing three-momentum $q=0$ has been reported in Ref. \cite{Ding}.
For practical applications, an extrapolation to finite $q$ values
is needed. A construction proposed in Ref. \cite{Rapp_3} and employed
in our calculation reads 
\begin{equation}
\begin{split}\frac{d^{4}R}{d^{4}q}=\frac{\alpha^{2}}{4\pi^{4}}f^{B}(q_{0},T)C_{{\rm EM}}\{1+\frac{2T}{q}\ln[\frac{1+x_{+}}{1+x_{-}}]\\
+2\pi\alpha_{s}\frac{T^{2}}{M^{2}}K\,F(M^{2})\ln(1+\frac{2.912q_{0}}{4\pi\alpha_{s}T})\frac{2q_{0}^{2}+M^{2}}{3q_{0}^{2}}\},\label{eq:rateQGP}
\end{split}
\end{equation}
where $\alpha$ is the electromagnetic coupling constant, $f^{B}(q_{0},T)$
the thermal Bose distribution, $C_{{\rm EM}}\equiv\sum_{q=u,d,s}e_{q}^{2}$,
$x_{\pm}=\exp[-(q_{0}\pm q)/2T]$. The quantity $\alpha_{s}$ is the
temperature-dependent strong coupling constant, $K$ a constant factor
(equal 2), and finally we have a form factor $F(M^{2})=\frac{\Lambda^{2}}{\Lambda^{2}+M^{2}}$
with $\Lambda=2T$.

\subsection{Rates from hadronic medium}

According to the vector meson dominance model \cite{Gounaris}, the
hadronic electromagnetic current operator is equal to the linear combination
of the known neutral vector meson field operators, most notably V=$\rho$,
$\omega$, $\phi$. This describes dilepton production successfully
\cite{Rapp_1}, where the dilepton emission rate via the vector meson
V is \cite{Vujanovic} 
\begin{equation}
\frac{d^{4}R_{{\rm V}}}{d^{4}q}=-\frac{\alpha^{2}m_{{\rm V}}^{4}}{\pi^{3}M^{2}g_{{\rm V}}^{2}}{\rm Im}D_{{\rm V}}\frac{1}{e^{\frac{q_{0}}{T}}-1},\label{eq:rateHG}
\end{equation}
with the coupling constant $g_{{\rm V}}$ determined by the measured
decay rate of vector-meson to dilepton production in vacuum, and $m_{{\rm V}}$
being the mass of the vector-meson. The imaginary part of the retarded
vector meson propagator is given as 
\begin{equation}
{\rm Im}D_{{\rm V}}=\frac{{\rm Im}\Pi_{{\rm V}}}{(M^{2}-m_{{\rm V}}^{2}-{\rm Re}\Pi_{{\rm V}})^{2}+({\rm Im}\Pi_{{\rm V}})^{2}},\label{eq:ImD}
\end{equation}
where $\Pi_{{\rm V}}$ is the self-energy of the vector meson V.

We consider V=$\rho$ and neglect the complexity of $\omega$ and
$\phi$, which seems consistent with the STAR dilepton data taking\cite{STAR_PRL}.
The $\rho$ meson self-energy 
\begin{equation}
\Pi_{\rho}=\Pi_{\rho}^{vac}+\sum_{a}\Pi_{\rho a}\:,\label{eq:selfE2terms}
\end{equation}
for the contribution from the vacuum and from $\rho$ meson scattering
with hadrons of type $a$ in the hadronic gas, respectively.

The vacuum part $\Pi_{\rho}^{vac}$ is obtained from Gounaris-Sakurai
formula as Refs \cite{Gounaris,Kapusta}, which could describle the
pion electromagnetic form factor well, as measured in $e^{+}e^{-}$
annihilation Refs \cite{Strauch}: 
\begin{equation}
\begin{split}{\rm Re}\Pi_{\rho}^{vac}=\frac{g_{\rho}^{2}M^{2}}{48\pi^{2}}[(1-\frac{4m_{\pi}^{2}}{M^{2}})^{\frac{3}{2}}\ln\left|\frac{1+\sqrt{1-\frac{4m_{\pi}^{2}}{M^{2}}}}{1-\sqrt{1-\frac{4m_{\pi}^{2}}{M^{2}}}}\right|\\
+8m_{\pi}^{2}(\frac{1}{M^{2}}-\frac{1}{m_{\rho}^{2}})-2(\frac{p_{0}}{\omega_{0}})^{3}\ln(\frac{\omega_{0}+p_{0}}{m_{\pi}})],\label{eq:ReSelfE-vac}
\end{split}
\end{equation}

\begin{equation}
{\rm Im}\Pi_{\rho}^{vac}=-\frac{g_{\rho}^{2}M^{2}}{48\pi}(1-\frac{4m_{\pi}^{2}}{M^{2}})^{\frac{3}{2}}.\label{eq:ImSelfE-vac}
\end{equation}
Here, $2\omega_{0}=m_{\rho}=2\sqrt{m_{\pi}^{2}+p_{0}^{2}}$. The vacuum
width is $\Gamma_{\rho}^{vac}=\frac{g_{\rho}^{2}}{48\pi}m_{\rho}(\frac{p_{0}}{\omega_{0}})^{3}$.

The interactive part $\Pi_{\rho a}(E,p)$ is obtained from $\rho$
scattering from hadron of type $a$ in the hadronic gas 
\begin{equation}
\Pi_{\rho a}(E,p)=-4\pi\int\frac{d^{3}k}{(2\pi)^{3}}n_{a}(\omega)\frac{\sqrt{s}}{\omega}f_{\rho a}^{c.m.}(s),\label{eq:SelfE-inter}
\end{equation}
where $f^{c.m.}$ is the forward scattering amplitude in the c.m.
system, $E$ and $p$ are the energy and momentum of the $\rho$ meson,
$\omega^{2}=m_{a}^{2}+k^{2}$. The most copious hadrons in the hadronic
gas such as $\pi$ mesons ($a=\pi$) and nucleons ($a=N$) are considered
in our calculation. The quantity $n_{a}$ is the Bose-Einstein occupation
number of $\pi$ mesons and the Fermi-Dirac occupation number for
nucleons, with temperature $T$ and baryon chemical potential $\mu_{B}$
of the thermal bath provided by EPOS3 mentioned above. According to
Eq.(\ref{eq:lorentztr1}-\ref{eq:lorentztr2}), the emitted dileptons
will get a Lorentz boost with the collective flow velocity offered
by EPOS3, so the interactive term of self-energy and the resulted
emission rate are calculated in the rest frame of the thermal bath.

The $\rho a$ forward scattering amplitude in the center-of-mass frame
could be written as 
\begin{equation}
f_{\rho a}^{c.m.}=(1-x)f_{{\rm Res}}+xf_{{\rm Reg}}+f_{{\rm Pom}},
\end{equation}
where the Pomeron term is dual to the background upon which the resonances
are superimposed, $x$ is a function that matches the low energy Breit-Wigner
resonances and high energy Reggons (dual to s-channel resonances)
smoothly: 
\begin{equation}
x=0.5(1+\tanh(\frac{E_{\rho}-m_{\rho}-E_{\Delta}}{0.3})),
\end{equation}
where $E_{\Delta}$ is 1GeV and 4GeV for $\rho\pi$ and $\rho N$
scattering, respectively\cite{Xu1}.

Regge and Pomeron term have the same form \cite{Eletsky}: 
\begin{equation}
f_{{\rm Reg/Pom}}=-\frac{q_{c.m.}}{4\pi s}\frac{1+e^{-i\pi\alpha}}{\sin\pi\alpha}s^{\alpha}r^{\rho a},
\end{equation}
where the intercept $\alpha$ and residue $r^{\rho N}$, $r^{\rho\pi}$
are 0.642, 28.59, 12.74 for Regges and 1.093, 11.88, 7.508 for Pomerons,
respectively (the units yield a cross section in mb with energy in
GeV) .

The resonance term is \cite{Eletsky}: 
\begin{equation}
f_{{\rm Res}}=\frac{1}{2q_{c.m.}}\sum_{R}W_{\rho a}^{R}\frac{\Gamma_{R\to\rho a}}{M_{R}-\sqrt{s}-\frac{1}{2}i\Gamma_{R}},
\end{equation}
which involves a sum over a series of Breit-Wigner resonances of mass
$M_{R}$ and total width $\Gamma_{R}$. The resonances R used in our
calculation are listed in table I for $a=\pi$ and table II for $a=N$,
with R's name, mass, decay width, branching ratio, spin, isospin and
relative angular momentum Ref. \cite{Eletsky,Xu1}.

\begin{table}[!hbp]
\protect\caption{Meson resonances $R$ for $\rho\pi$ processes.}

\begin{tabular}{|c|c|c|c|c|c|c|}
\hline 
Name(R)  & Mass($M_{R}$)  & $\Gamma$  & BR  & S  & IS  & L \tabularnewline
\hline 
$\phi$(1020)  & 1.020  & 0.0045  & 0.13  & 1  & 0  & 1 \tabularnewline
\hline 
$h_{1}$(1170)  & 1.170  & 0.36  & 1  & 1  & 0  & 0 \tabularnewline
\hline 
$a_{1}$(1260)  & 1.230  & 0.40  & 0.68  & 1  & 1  & 0 \tabularnewline
\hline 
$\pi$(1300)  & 1.300  & 0.40  & 0.32  & 0  & 1  & 1 \tabularnewline
\hline 
$a_{2}$(1320)  & 1.318  & 0.107  & 0.70  & 2  & 1  & 2 \tabularnewline
\hline 
$\omega$(1420)  & 1.419  & 0.174  & 1  & 1  & 0  & 1 \tabularnewline
\hline 
\end{tabular}
\end{table}

\begin{table}[!hbp]
\protect\caption{Baryon resonances $R$ for $\rho N$ processes.}

\begin{tabular}{|c|c|c|c|c|c|c|}
\hline 
Name(R)  & Mass($M_{R}$)  & $\Gamma$  & BR  & S  & IS  & L \tabularnewline
\hline 
N(2090)  & 1.928  & 0.414  & 0.49  & 0.5  & 0.5  & 0 \tabularnewline
\hline 
N(1700)  & 1.737  & 0.249  & 0.13  & 1.5  & 0.5  & 0 \tabularnewline
\hline 
N(2080)  & 1.804  & 0.447  & 0.26  & 1.5  & 0.5  & 0 \tabularnewline
\hline 
N(2190)  & 2.127  & 0.547  & 0.29  & 3.5  & 0.5  & 2 \tabularnewline
\hline 
N(2100)  & 1.885  & 0.113  & 0.27  & 0.5  & 0.5  & 1 \tabularnewline
\hline 
N(1720)  & 1.717  & 0.383  & 0.87  & 1.5  & 0.5  & 1 \tabularnewline
\hline 
N(1900)  & 1.879  & 0.498  & 0.44  & 1.5  & 0.5  & 1 \tabularnewline
\hline 
N(2000)  & 1.903  & 0.494  & 0.60  & 2.5  & 0.5  & 1 \tabularnewline
\hline 
$\Delta$(1900)  & 1.920  & 0.263  & 0.38  & 0.5  & 1.5  & 0 \tabularnewline
\hline 
$\Delta$(1700)  & 1.762  & 0.599  & 0.08  & 1.5  & 1.5  & 0 \tabularnewline
\hline 
$\Delta$(1940)  & 2.057  & 0.460  & 0.35  & 1.5  & 1.5  & 0 \tabularnewline
\hline 
$\Delta$(2000)  & 1.752  & 0.251  & 0.22  & 2.5  & 1.5  & 1 \tabularnewline
\hline 
$\Delta$(1905)  & 1.881  & 0.327  & 0.86  & 2.5  & 1.5  & 1 \tabularnewline
\hline 
N(1520)  & 1.520  & 0.124  & 0.008  & 1.5  & 0.5  & 0 \tabularnewline
\hline 
$\Delta$(1232)  & 1.232  & 0.118  & 0.006  & 1.5  & 1.5  & 1 \tabularnewline
\hline 
\end{tabular}
\end{table}

\begin{figure}
\includegraphics[scale=0.4]{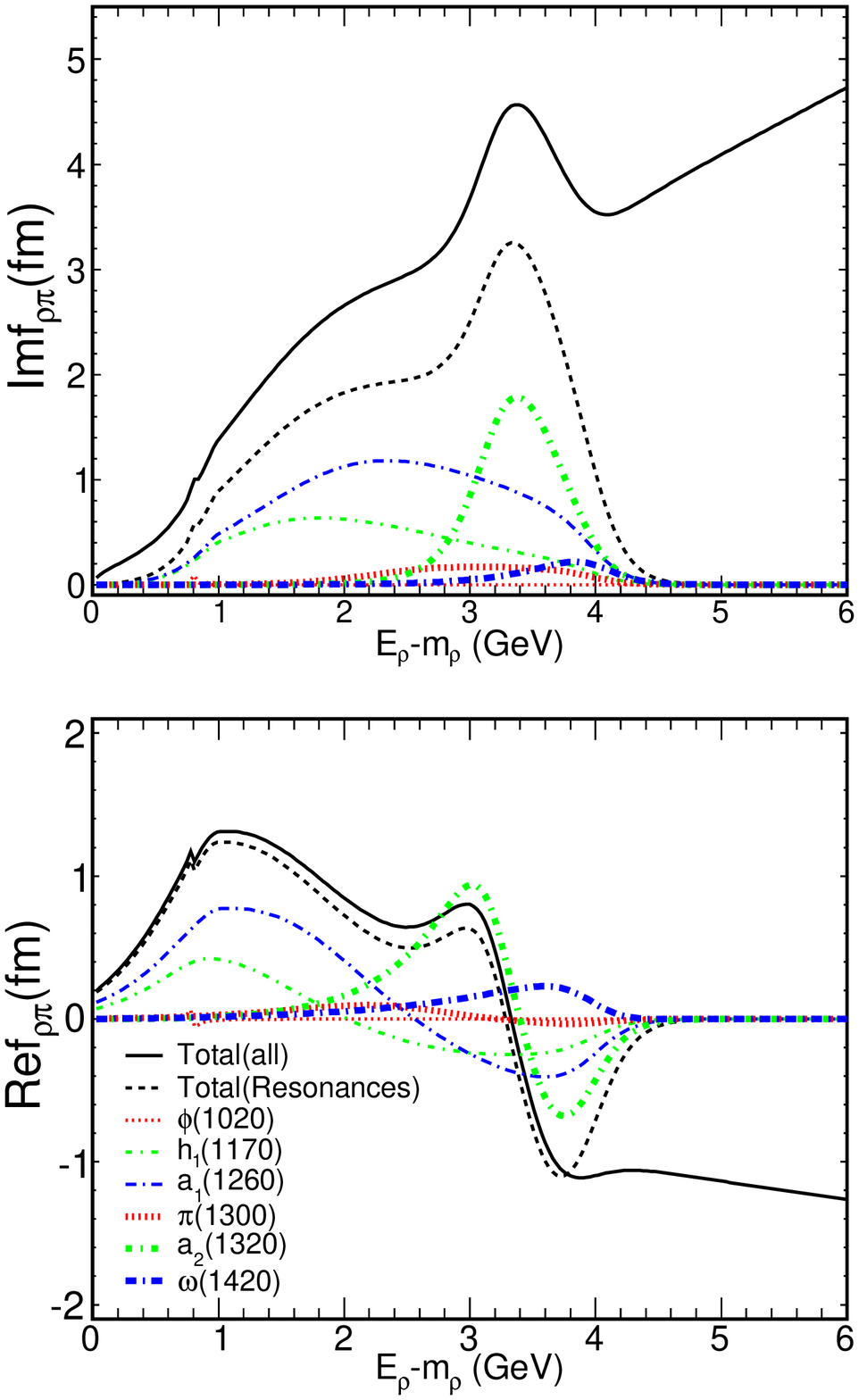}

\protect\caption{\label{fig:frhopi} (Color Online) The individual and total amplitude
for $\rho\pi\rightarrow R$ scattering (imaginary part: upper panel,
real part: lower panel).}
\end{figure}

\begin{figure}
\includegraphics[scale=0.4]{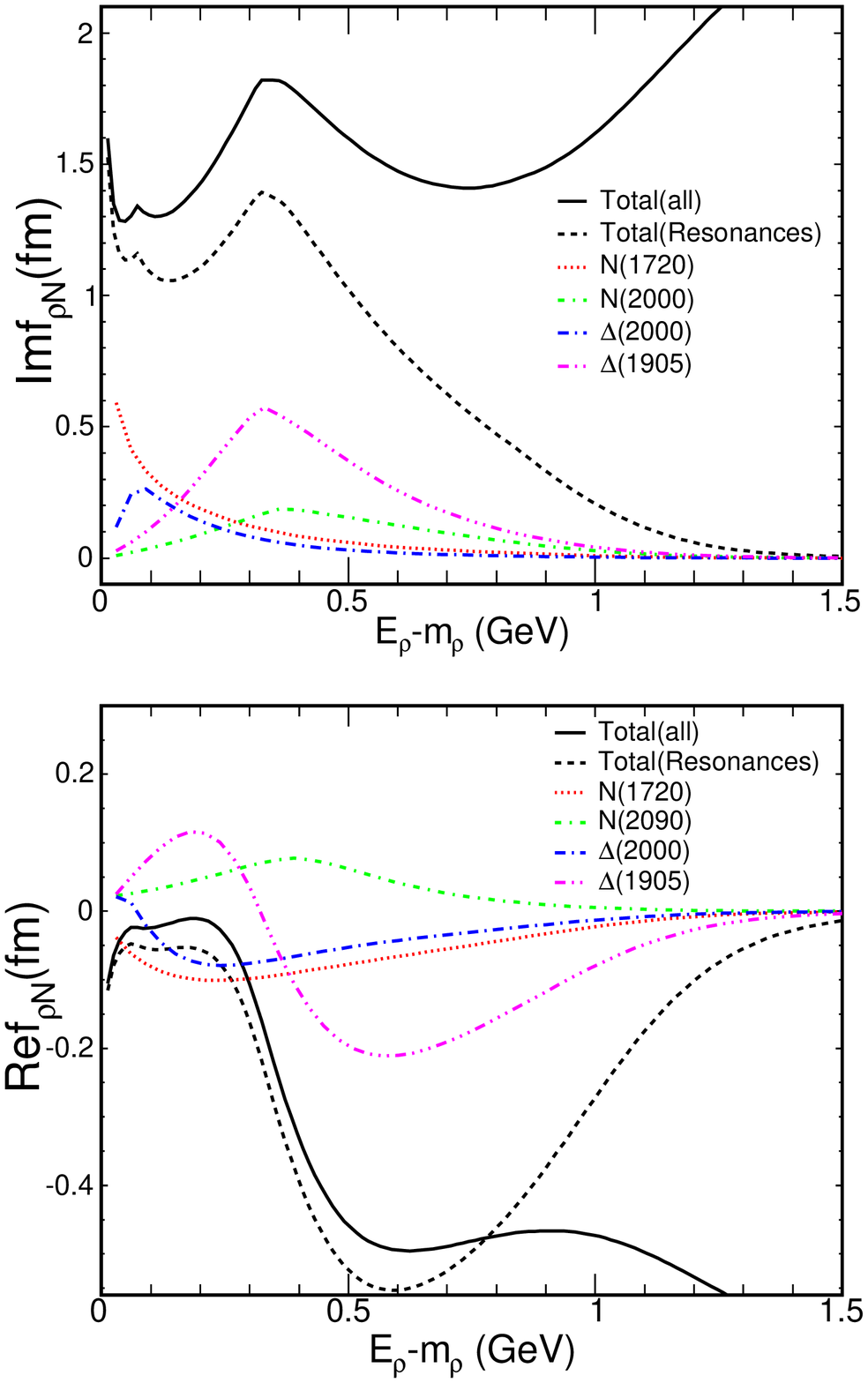}

\protect\caption{\label{fig:frhoN} (Color Online) The relevant individual and total
amplitude for $\rho N\rightarrow R$ scattering (imaginary part: upper
panel, real part: lower panel) .}
\end{figure}

The c.m. amplitude $f^{c.m.}$ and the scattering amplitude in the
rest frame of $a$, $f_{\rho a}$, are related by 
\begin{equation}
\sqrt{s}f_{\rho a}^{c.m.}(s)=m_{a}f_{\rho a}(E_{\rho}),\label{eq:frhoa}
\end{equation}
with $s=m_{\rho}^{2}+m_{a}^{2}+2E_{\rho}m_{a}$. The latter is plotted
in Fig.\ref{fig:frhopi} and Fig.\ref{fig:frhoN}, for $\rho\pi$
and $\rho N$ scattering, respectively. The total contribution (Resonances
+ Pomeron background + Regge) in both are shown as black solid lines,
and both are consistent with Ref.~\cite{Eletsky}. Each individual
process $\rho\pi\rightarrow R$ is also shown in Fig.\ref{fig:frhopi}.
Their sum is shown as black dashed line. We can see that the contribution
of resonances dominants the low energy scattering, and the dominant
channels are $R=a_{1}(1260)$ and $h_{1}(1170)$ due to their large
branch ratios. In Fig.\ref{fig:frhoN}, the relevant individual channels
$\rho N\rightarrow R$ are shown. The black dashed line is the total
baryon resonances contribution, among which $R=\Delta(1905),N(2000)$
and $N(1720)$ dominate both the imaginary and the real part of the
the amplitude for $\rho N$ scattering.

\begin{figure}
\includegraphics[scale=0.4]{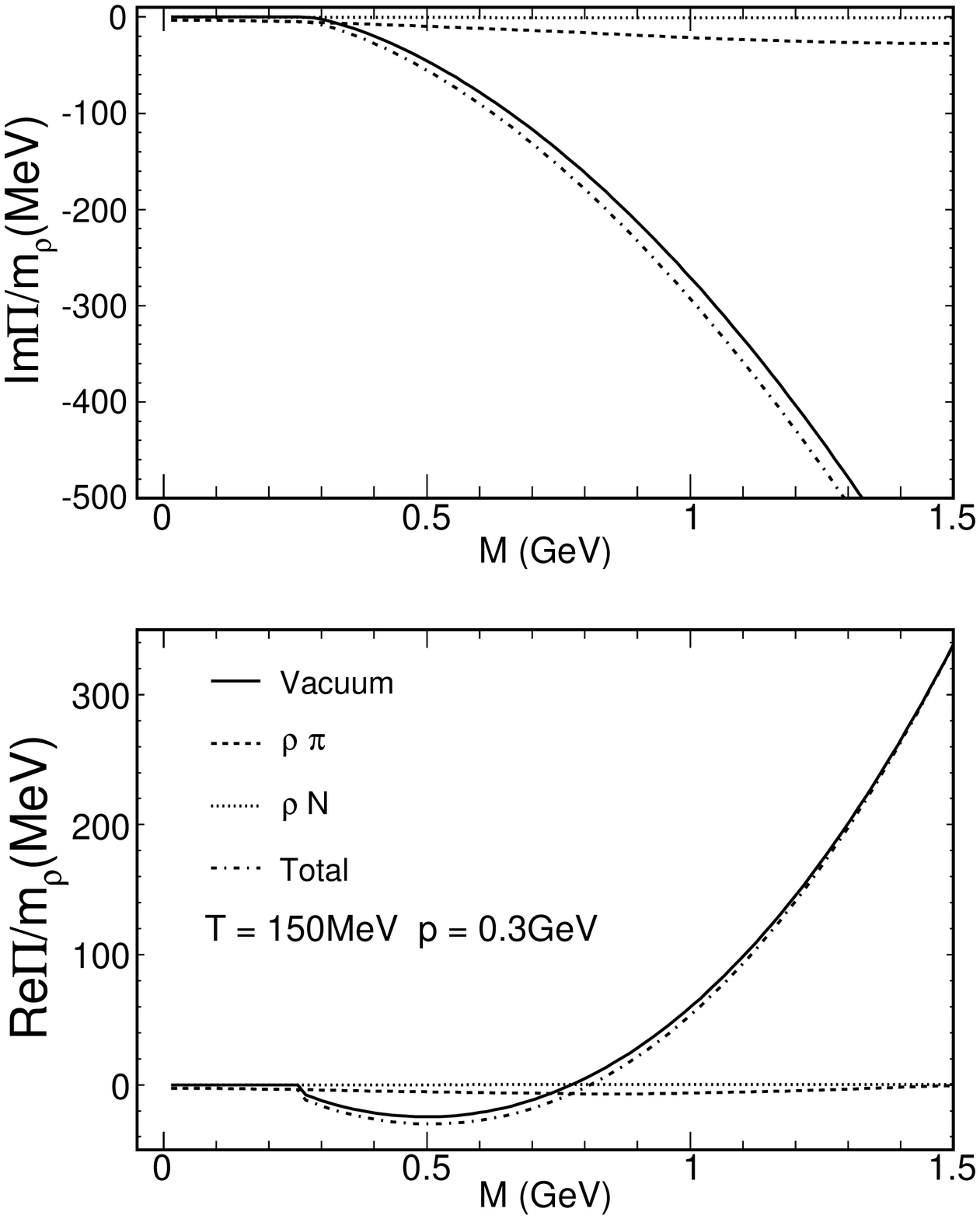}

\protect\caption{\label{fig:selfenergy1} The self-energy of $\rho$-meson .}
\end{figure}

Putting the forward scattering amplitutes together, we get the interactive
term of self-energy of $\rho$-meson. The imgaginary and real parts
are shown in Fig.\ref{fig:selfenergy1}, for a fixed $\rho$-meson
three-momentum $q=0.3$~GeV/c at $T=150$~MeV, with $\rho\pi$ scattering
dashed lines and $\rho N$ scattering dotted lines, respectively.
The contribution from vacuum is plotted as solid lines. The total
one is plotted as dashed-dotted lines. The vacuum contributes when
$M>2m_{\pi}$, then dominates over a wide region. The in-medium effect
from interactive term seems rather small.

\begin{figure}
\includegraphics[scale=0.3]{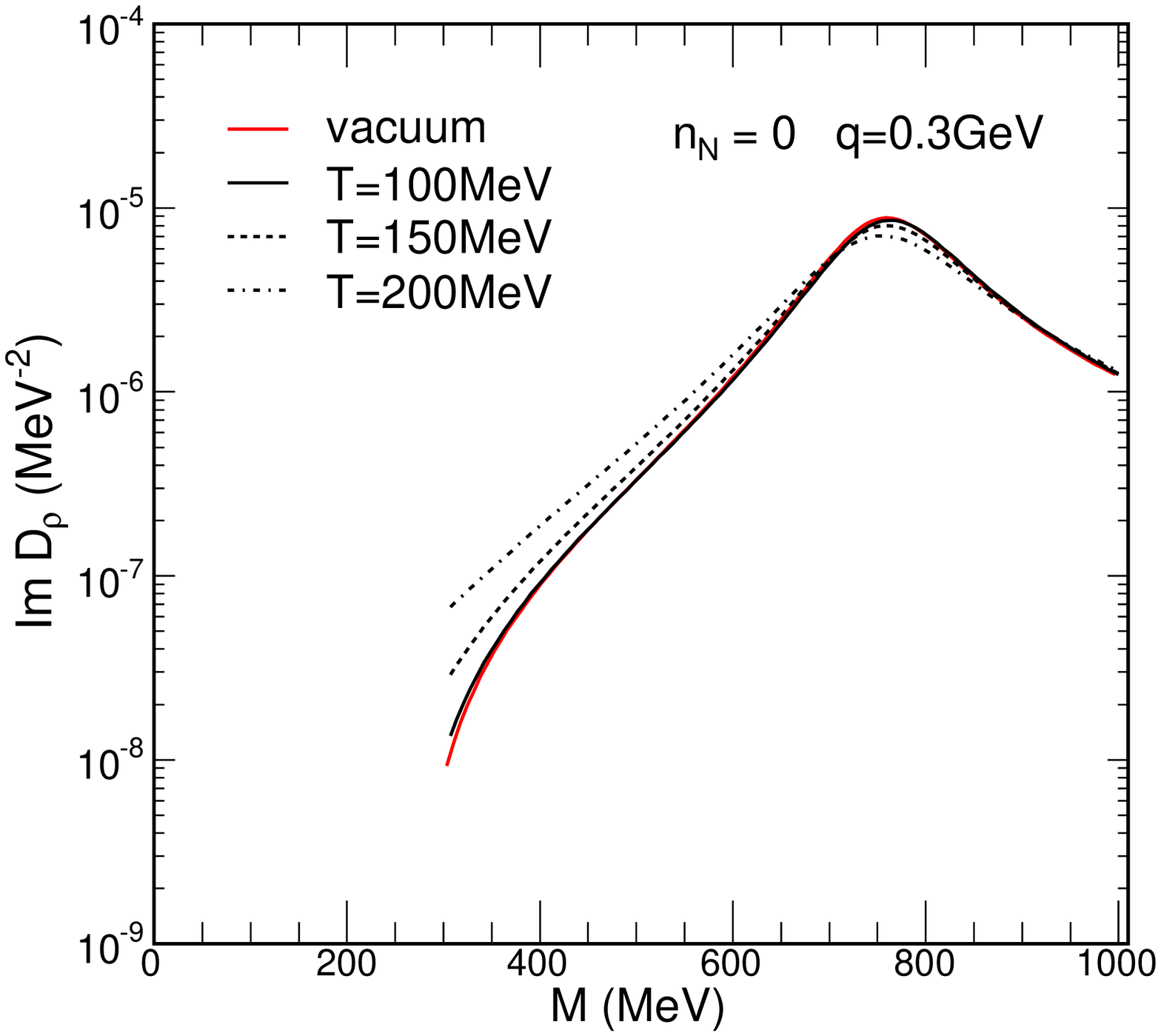}

\protect\caption{\label{fig:rhopropagator1} (Color Online) The imaginary part of the
$\rho$ meson propagators as a function of invariant mass for a momentum
of $300MeV/c$ and a nucleon density of $n_{N}=0$. Results are shown
for vacuum(red solid line) and three temperatures, $T=100MeV/c$ (black
solid line), $T=150MeV/c$ (black dashed line), $T=200MeV/c$ (black
dashed-dotted line).}
\end{figure}

Yet, the interaction with the medium gives a broadening effect to
the spectral density, $\rho(q)=-{\rm Im}D_{\rho}(q)/\pi$. In Fig.\ref{fig:rhopropagator1},
${\rm Im}D_{\rho}$ is plotted as a function of invariant mass $M$
for a $\rho$-meson momentum of $q=0.3$~GeV/c, in a thermal hadronic
gas of temperature T=100, 150 and 200MeV, respectively. The vacuum
contribution by itself is plotted as a red solid line. With the increase
of medium temperature, the $\rho$-meson scattering with hadron $a$
in the medium broadens the spectral density more and more, but the
peak remains close to $\rho$-meson mass. (Note: T=200MeV is only
to visualize the broadening effect, though it is too high for hadronic
gas. )

\begin{figure}
\includegraphics[scale=0.3]{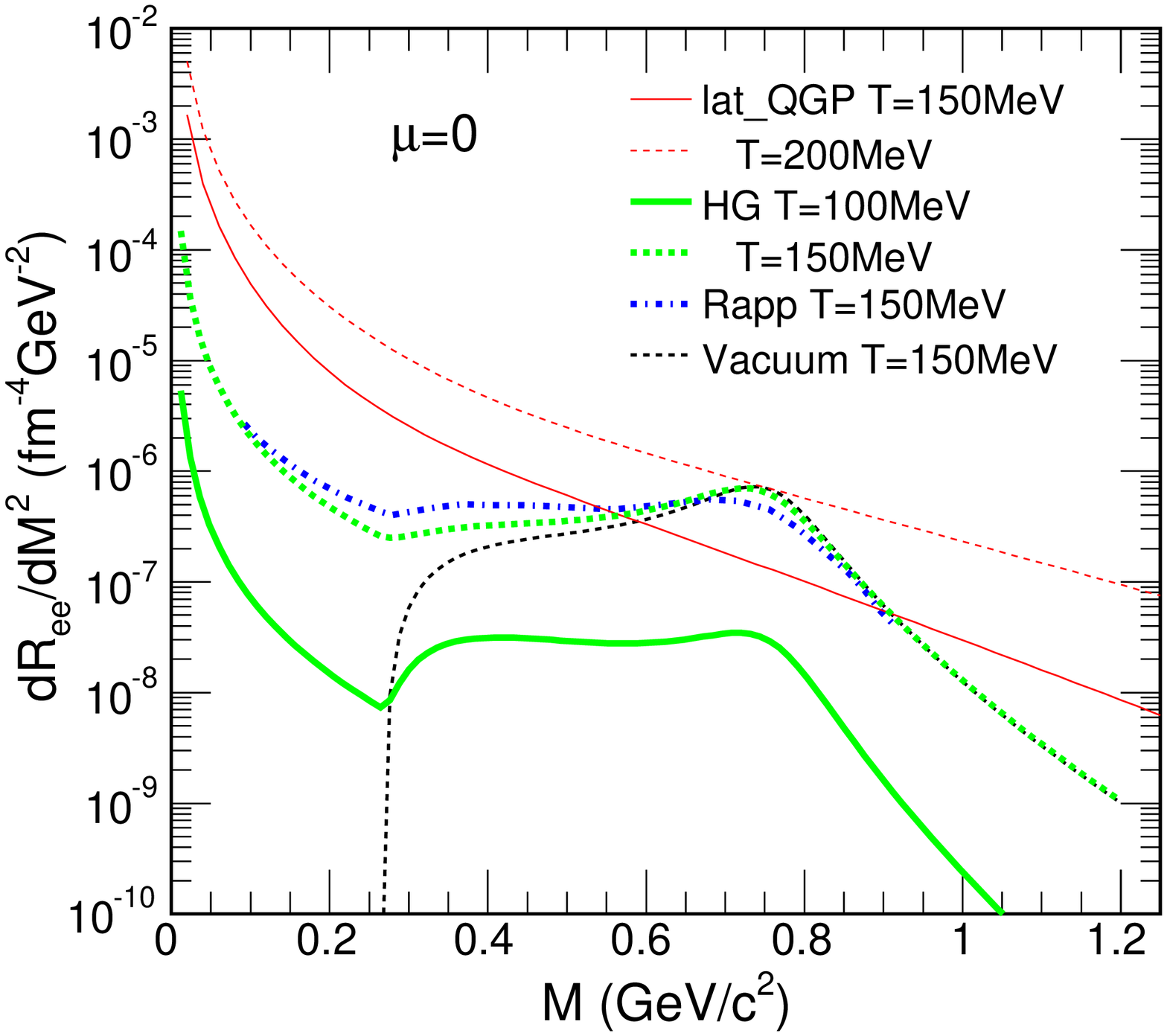}

\protect\caption{\label{fig:rate1} (Color Online) Emission rates of dielectrons from
QGP phase (red thin lines) at temperature: 150MeV (solid line), 200MeV
(dashed line), and HG phase (green thick lines) at temperature: 100MeV
(solid line), 150MeV (dashed line). The vacuum contribution to HG
rate at 150MeV is shown as black dashed line. HG rate from effective
interaction Lagrangians \cite{Rapp_4} at 150MeV is shown as blue
dashed-dotted line.}
\end{figure}

The emission rates at different phases are now obtained via Eq.(\ref{eq:rateQGP}-\ref{eq:rateHG}).
In Fig.\ref{fig:rate1}, the emission rate from the QGP phase is plotted
as red thin lines (dotted: 200MeV, solid: 150MeV), from the HG phase
as green thick lines (dotted: 150MeV, solid: 100MeV). A higher temperature
makes a stronger di-electron emission. Nevertheless the HG phase provides
a pronounced peak around the $\rho$-meson mass 775MeV which exceeds
the QGP contribution. This peak is essentially due to the vacuum term
in the self-energy (black dashed line). The emission rate from HG
phase at 150MeV obtained from effective interaction Lagrangians \cite{Rapp_4}
is also shown as a blue dotted dashed line. It is interesting to see
the two rates are quite close to each other, though different approaches
and channels are considered.

We should mention here that the above-mentioned emission rates work
typically for ideal hydrodynamics. For viscous hydrodynamics such
as EPOS3, a viscous correction is needed. The effect of shear viscosity
to the spectra and elliptic flow of dileptons in QGP phase and HG
phase has been investigated~\cite{Vujanovic:2012nq}. Similar work
has been done for direct photons~\cite{Shen:2013vja}. The elliptic
flow of dileptons seems more sensitive to viscous effect than their
spectra, however, it remains a modest effect.

\subsection{Elliptic flow $v_{2}$ and Triangular flow $v_{3}$}

The elliptic flow $v_{2}$ and the triangular one $v_{3}$ of thermal
dielectrons are calculated in a similar way as in Refs \cite{Fu}.
The azimuthal angle dependence of the invariant mass spectrum of dielectrons
for a given event can be decomposed into harmonics of the azimuthal
angle $\phi$ as 
\begin{equation}
\frac{dN}{d\phi} \sim\frac{1}{2\pi}[1+2v_{2}\cos2(\phi-\psi_{2})+2v_{3}\cos3(\phi-\psi_{3})+\cdots],\label{eq:dndphi}
\end{equation}
where $v_{2}$($v_{n}$) is the elliptic flow (higher order harmonics),
and $\psi_{n}$ is the $n$th-order event plane angle. Obviously,
$v_{n}$ and $\psi_{n}$ depend on the dielectron's invariant mass
$M$ and vary event by event. From Eq.~(\ref{eq:dndphi}), one can
easily get 
\begin{equation}
v_{e,n}\cos(n\psi_{e,n})=\frac{\int_{0}^{2\pi}\cos(n\phi)\frac{dN}{d\phi} |_e  d\phi}{\int_{0}^{2\pi}\frac{dN}{d\phi} |_e  d\phi}
\end{equation}
\begin{equation}
v_{e,n}\sin(n\psi_{e,n})=\frac{\int_{0}^{2\pi}\sin(n\phi)\frac{dN}{d\phi} |_e  d\phi}{\int_{0}^{2\pi}\frac{dN}{d\phi} |_e d\phi},
\end{equation}
with the subscript $e$ added to emphasis variables for a single event. 
Let's note their right sides as $<\cos n\phi>_e$ and $<\sin n\phi>_e$,
respectively. Then, for each event, the harmonics and reaction plane
of order $n$ can be obtained as 
\begin{equation}
v_{e,n}=\sqrt{ {<\cos(n\phi)>_e}^{2}+{<\sin(n\phi)>_e}^{2}}
\end{equation}
and 
\begin{equation}
\psi_{e,n}=\frac{1}{n}\arctan\frac{<\sin(n\phi)>_e}{<\cos(n\phi)>_e}.
\end{equation}
The elliptic flow $v_{2}$ and higher order harmonics $v_{n}$ of
the sample is obtained via event average
\begin{equation}
v_{n}= \sum_{e=1}^N \frac {v_{e,n}}{N}.      
\end{equation}
 This is equivalent to the definition in \cite{Gale:2012rq}.

\section{RESULTS}

The results of thermal dielectrons calculated based on EPOS with initial
time $\tau_{0}=0.35fm/c$ will be shown in the following.

\begin{figure*}
\includegraphics[scale=0.7]{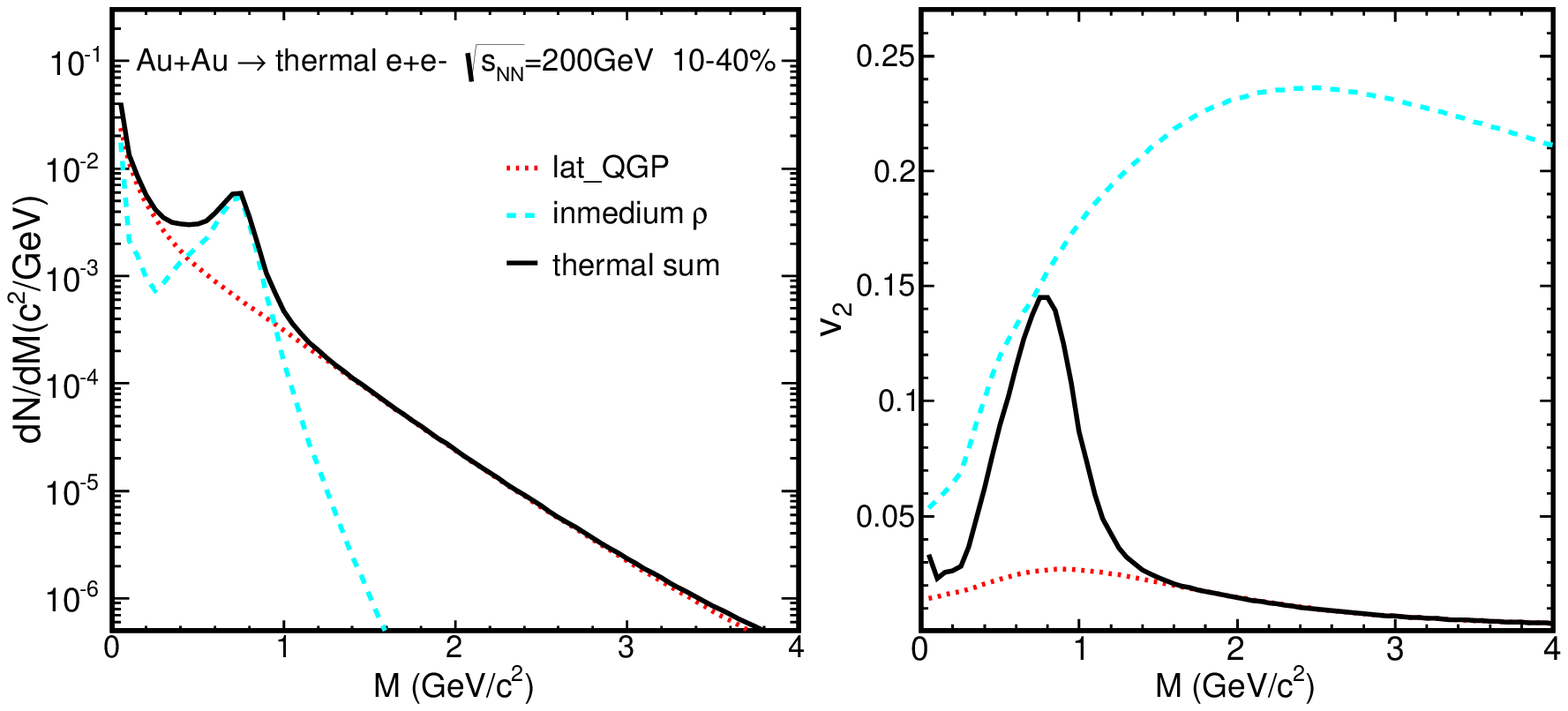}

\protect\caption{\label{fig:dndmandv2_10-40} (Color Online) Thermal di-electrons (black
solid lines) from AuAu collisions at $\sqrt{s_{NN}}$=200~GeV with
centrality 10-40\% are decomposed into di-electrons from QGP phase
(red dotted lines) and hadronic phase (blue dashed lines). Left panel:
the invariant mass spectra. Right panel: the elliptic flow. All calculations
are based on EPOS3+DiL1.}
\end{figure*}

In Fig.\ref{fig:dndmandv2_10-40} is shown the invariant mass spectrum
(left panel) and elliptic flow (right panel) of thermal di-electrons
at rapidity $y=0$ from AuAu collisions at $\sqrt{s_{NN}}=200$~GeV
with centrality 10-40\%. The thermal results (black solid lines) have
been decomposed into two contributions, coming from the QGP phase
(red dotted lines) and the hadronic phase (blue dashed lines).

From the left plot of Fig.\ref{fig:dndmandv2_10-40}, it is clear
that the QGP contribution dominates the thermal spectra at $M$>1GeV/$c^{2}$
region, as in the above figure on emission rates. And the peak from
the contribution of the hadronic phase indeed remains around the $\rho$
mass after the space-time integral of the emission rate. Thus, in
the total thermal spectrum, a pronounced peak still exists.

On the right panel of Fig.\ref{fig:dndmandv2_10-40}, the elliptic
flow of dileptons from the QGP phase (dotted line) and from the HG
(dashed line) are shown. The elliptic flow of the latter phase is
much larger due to a longer development of radial flow. The result
of thermal dileptons (solid line), the average of the two flows weighted
by their yields in each phase, ranges between the upper bound (values
in hadronic phase) and lower bound (values in QGP phase), and is close
to the corresponding bound of the dominating phase. Therefore, no
evident peak for the elliptic flows of dileptons from either phase
still produces a peak of thermal dileptons around the $\rho$-meson
mass.

A similar phase competition in the invariant mass spectrum and $v_{2}$
occurs at other collision centralities, ie, 0-10\% and 40-80\%.

After investigating the thermal contribution, we would like to compare
with data. Thus a study of non-thermal contribution is needed. EPOS3
is almost ready for this goal. For this moment, the employment of
STAR cocktail data can also make an independent check of the thermal
emission. STAR cocktail contribution includes all the non-thermal
contributions, such as contributions from Drell-Yan process and decay
of long lived hadrons, such as light mesons, D-mesons and so on. Based
on the measurement of those hadrons, Tsallis fitting and decay simulation,
STAR offers the invariant mass spectra of the cocktail contribution
\cite{STAR2}.

\begin{figure}
\includegraphics[scale=0.4]{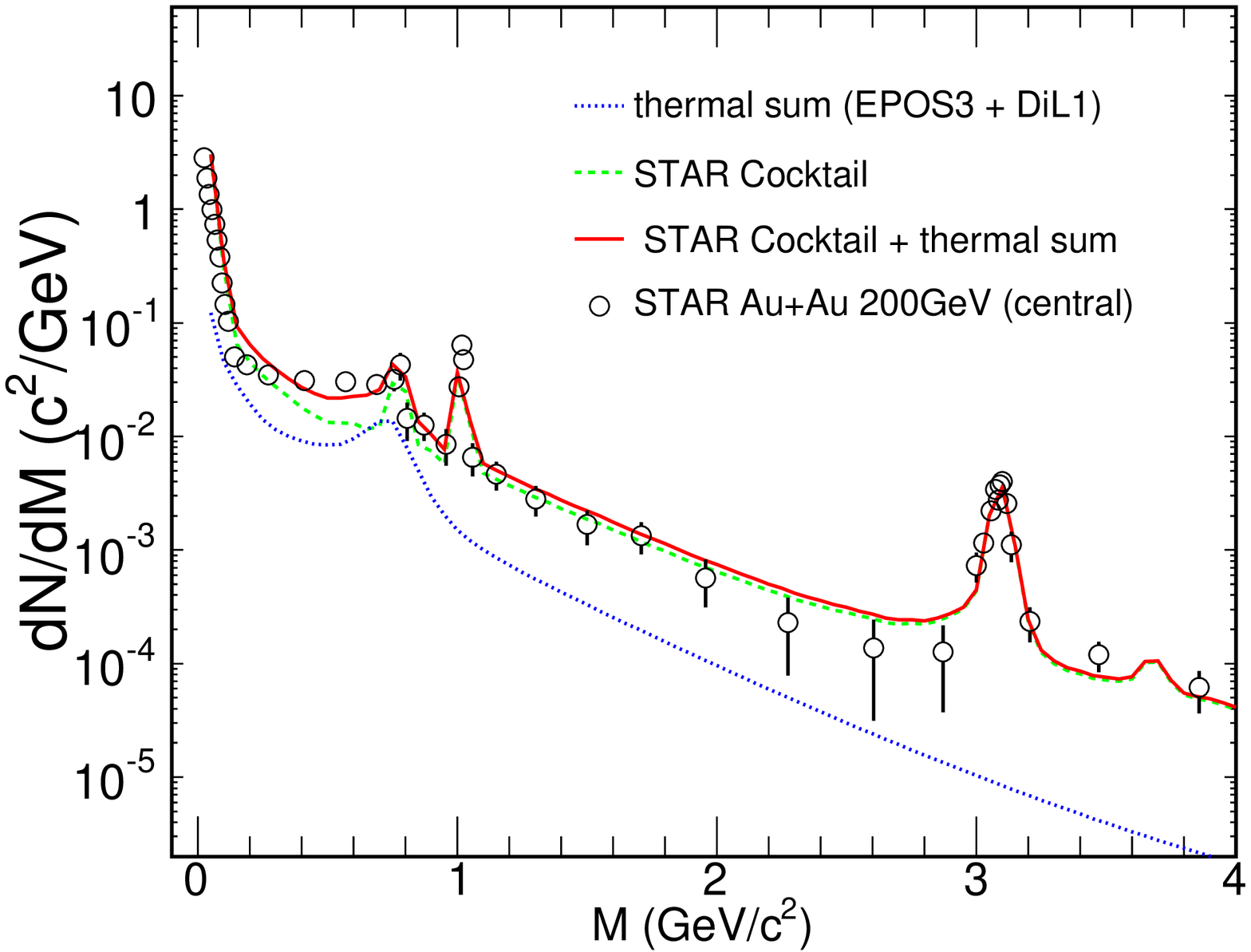}

\protect\caption{\label{fig:dndm0-10} (Color Online) Invariant mass spectra of dielectrons
from Au+Au collisions at $\sqrt{s_{NN}}=200$~GeV for centrality
0-10\%. The STAR data\cite{STAR_PRL} (empty dots) are compared to
the total contribution (solid line), which is the sum of the EPOS3+DiL1
thermal contribution (dotted line) and STAR cocktail (dashed line). }
\end{figure}

In Fig.\ref{fig:dndm0-10}, we show the EPOS3+DiL1 thermal contribution
(dotted line) and the STAR cocktail (dashed line) from 0-10\% Au+Au
central collisions at $\sqrt{s_{NN}}=200$~GeV. Their sum as the
invariant mass spectrum of dielectrons (solid line) is also compared
to STAR data~\cite{STAR_PRL} (empty dots).

\begin{figure*}
\includegraphics[scale=0.8]{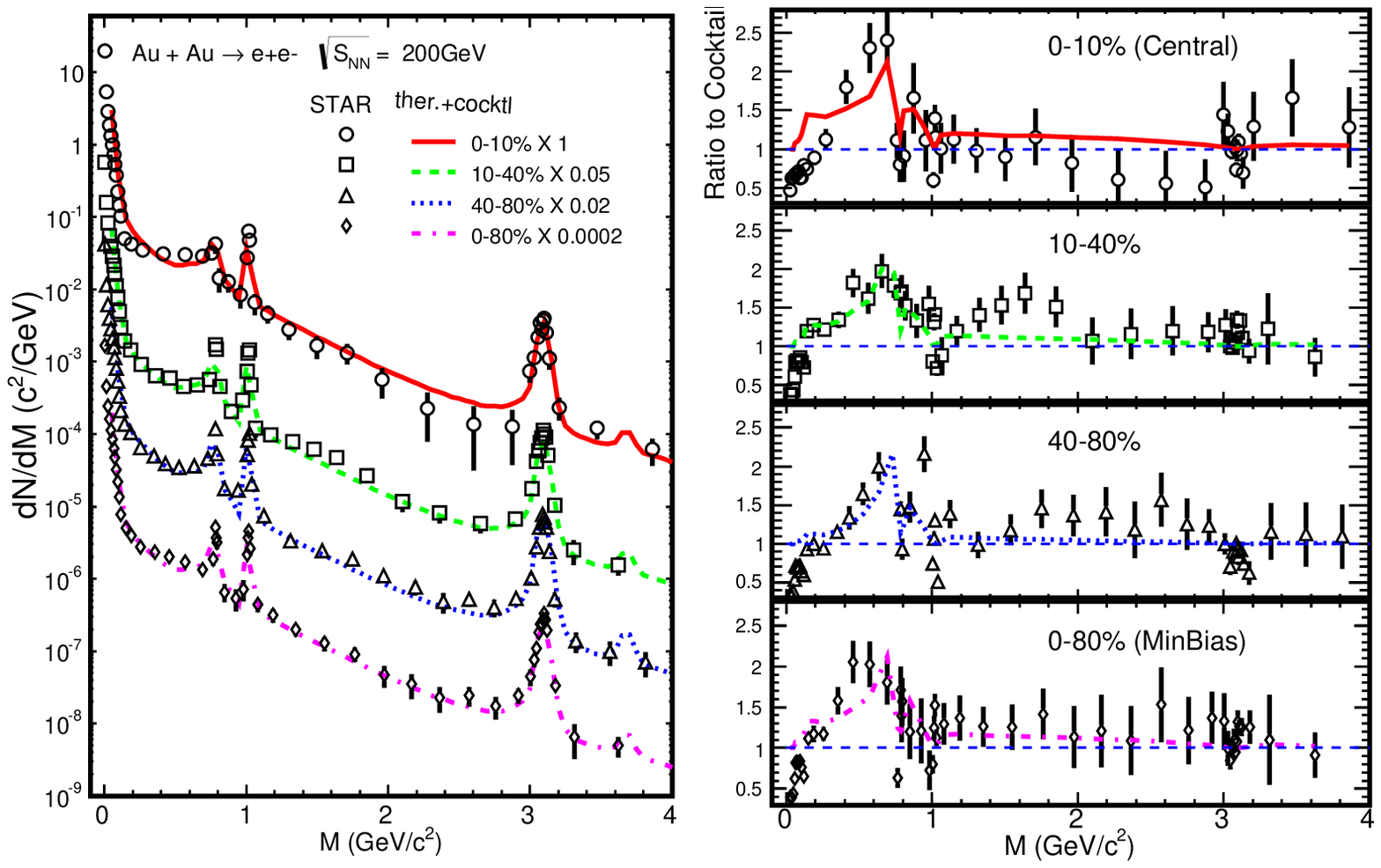}

\protect\caption{\label{fig:dndmandratio} (Color Online) Left panel: Invariant mass
spectra of dielectrons from EPOS3+DiL1 compared to STAR data \cite{STAR_PRL,STAR1},
for centrality 0-10\% (central), 10-40\%, 40-80\%, 0-80\% (MinBias)
(from top to bottom). Right panel: Ratios of STAR dilepton invariant
mass spectra and our calculations to cocktail, for centrality 0-10\%
(central), 10-40\%, 40-80\%, 0-80\% (MinBias) (from top to bottom). }
\end{figure*}

In Fig.\ref{fig:dndmandratio}(left panel) the invariant mass spectra
of dielectrons from Au+Au collisions at $\sqrt{s_{NN}}=200$~GeV
for different centralities from EPOS3+DiL1 are compared to STAR data
\cite{STAR_PRL,STAR1}, from top to bottom: 0-10\% (central), 10-40\%,
40-80\%, and 0-80\% (MinBias). For better visibility, the latter three
results are multiplied by factors: 0.05, 0.02, 0.0002, respectively.
The right panel shows the ratios of measured data and our calculations
to the cocktail at different centralities. This precise comparison
shows a deviation from the experimental data in low mass region for
the first centrality. The calculated results describe the data well
for centralities of 10-40\% and 40-80\% better than 0-10\%.
\begin{figure}
\includegraphics[scale=0.4]{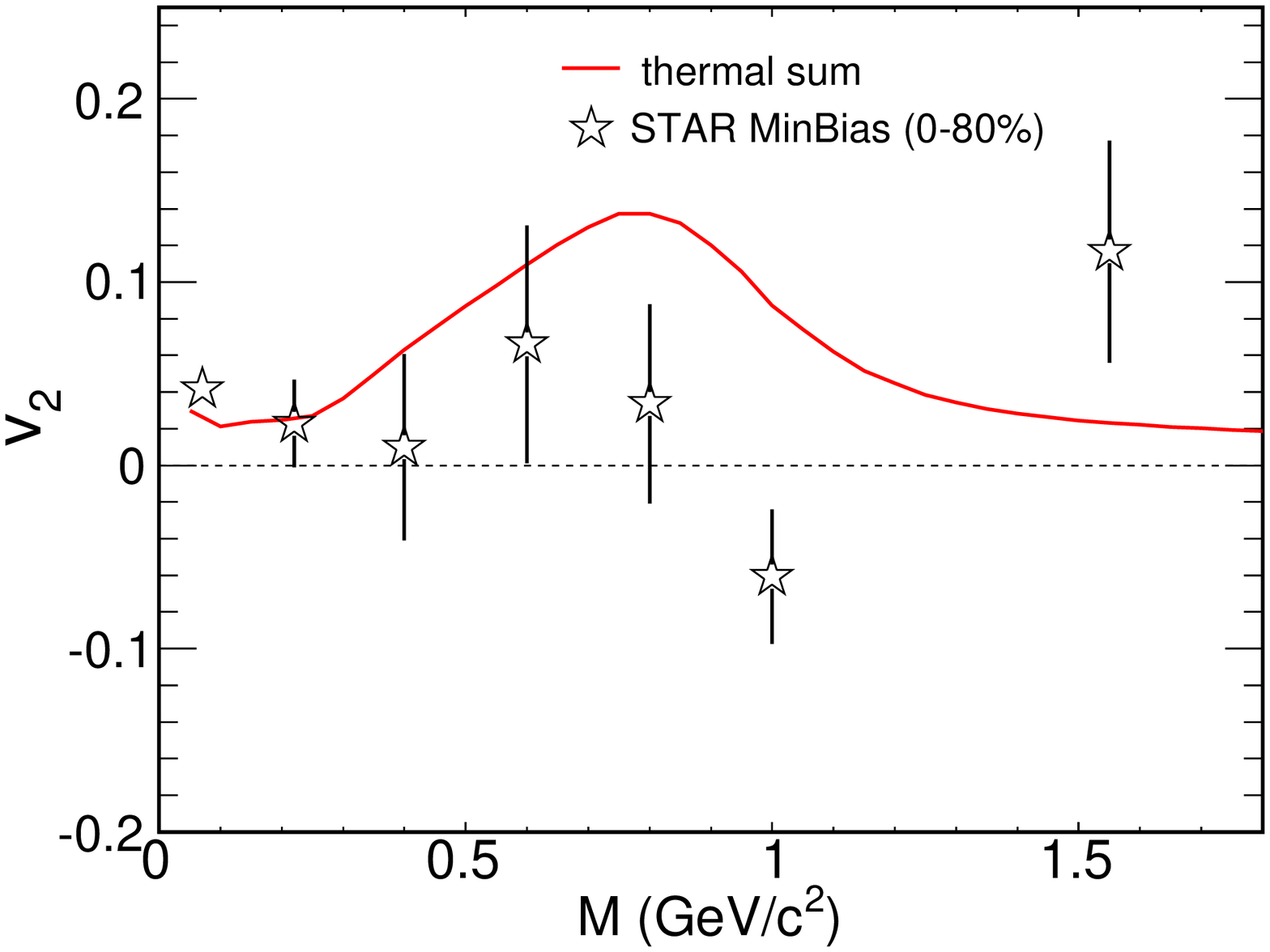}

\protect\protect\protect\protect\caption{\label{fig:v2MB} (Color Online) Thermal elliptic flow of dielectrons
from $\sqrt{s_{NN}}=200$~GeV Au+Au minimum bias collisions and comparison
with STAR data \cite{STAR_PRC}.}
\label{v2MB} 
\end{figure}

In Fig.\ref{fig:v2MB}, the red solid line is our calculated elliptic
flow of thermal di-electrons from Au+Au minimum bias collisions (0-80\%
centrality) at $\sqrt{s_{NN}}=200$~GeV. As a reference, we also
show STAR data \cite{STAR_PRC} (stars) of the elliptic flow of di-electrons
(from all sources, of course). The calculated elliptic flow of thermal
dielectrons is comparable, in fact even larger than data for all di-electrons.
The cocktail contribution will inherit a certain elliptic flow because
hadrons carry elliptic flow before decaying into di-electrons. Most
models predict it to be lower than the STAR reference \cite{Rapp_3,Vujanovic,Xu2},
whereas our result based on EPOS3 is larger.

To make a complete collection of our theoretical results, we present
the centrality dependence and the $n$ dependence of the flow harmonics
$v_{n}$ of dileptons in the following. 
\begin{figure*}
\includegraphics[scale=0.8]{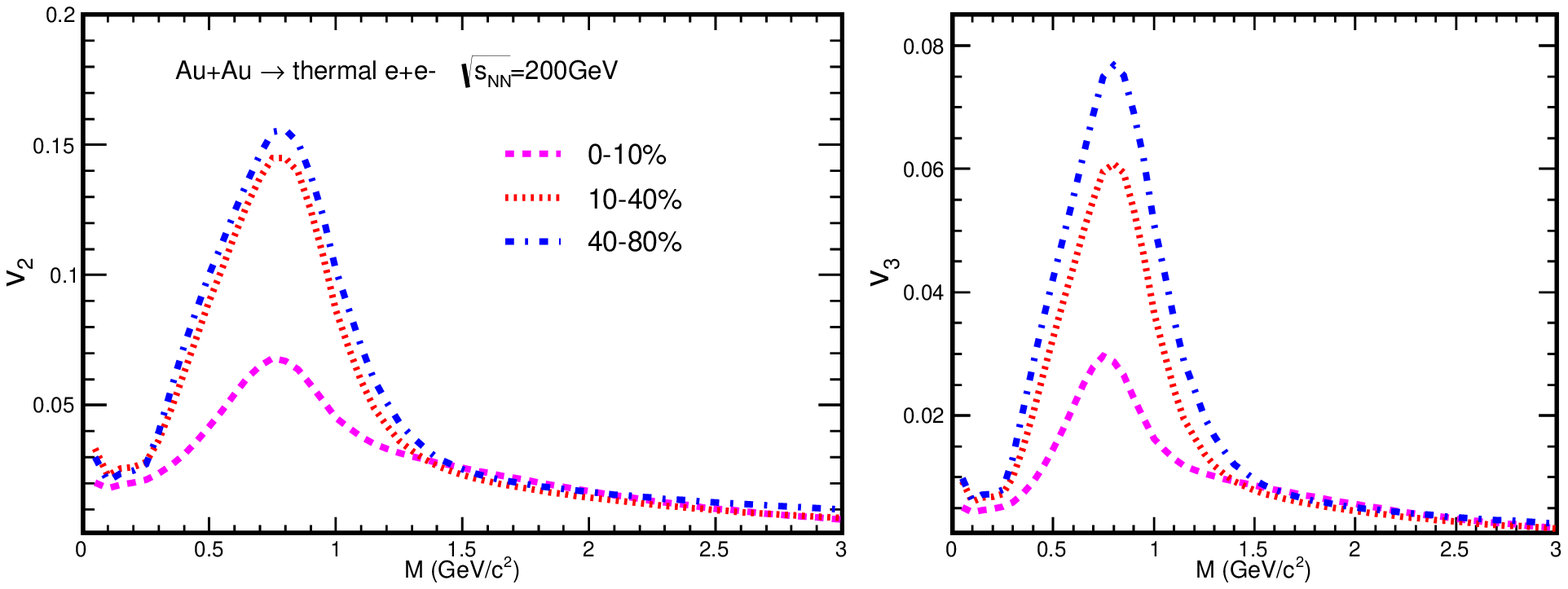}

\protect\caption{\label{fig:v2andv3} (Color Online) The elliptic flow (left) and trianger
flow (right) of thermal dielectrons from AuAu collisions at $\sqrt{s_{NN}}=200$~GeV.
0-10\%(pink dashed line), 10-40\%(red dotted line), 40-80\%(dark blue
dashed-dotted line).}
\end{figure*}

In Fig.\ref{fig:v2andv3} is shown the elliptic flow of thermal dielectrons
from AuAu collisions at $\sqrt{s_{NN}}=200$~GeV for different centralities:
0-10\%(pink dashed line), 10-40\%(red dotted line), 40-80\%(dark blue
dashed-dotted line). The centrality dependence of high order $n$
is also investigated. A strong centrality dependence, from central
to peripheral collisions, occurs not only to the elliptic flow but
also to higher order, $n=3,4,5$.

\begin{figure}
\includegraphics[scale=0.4]{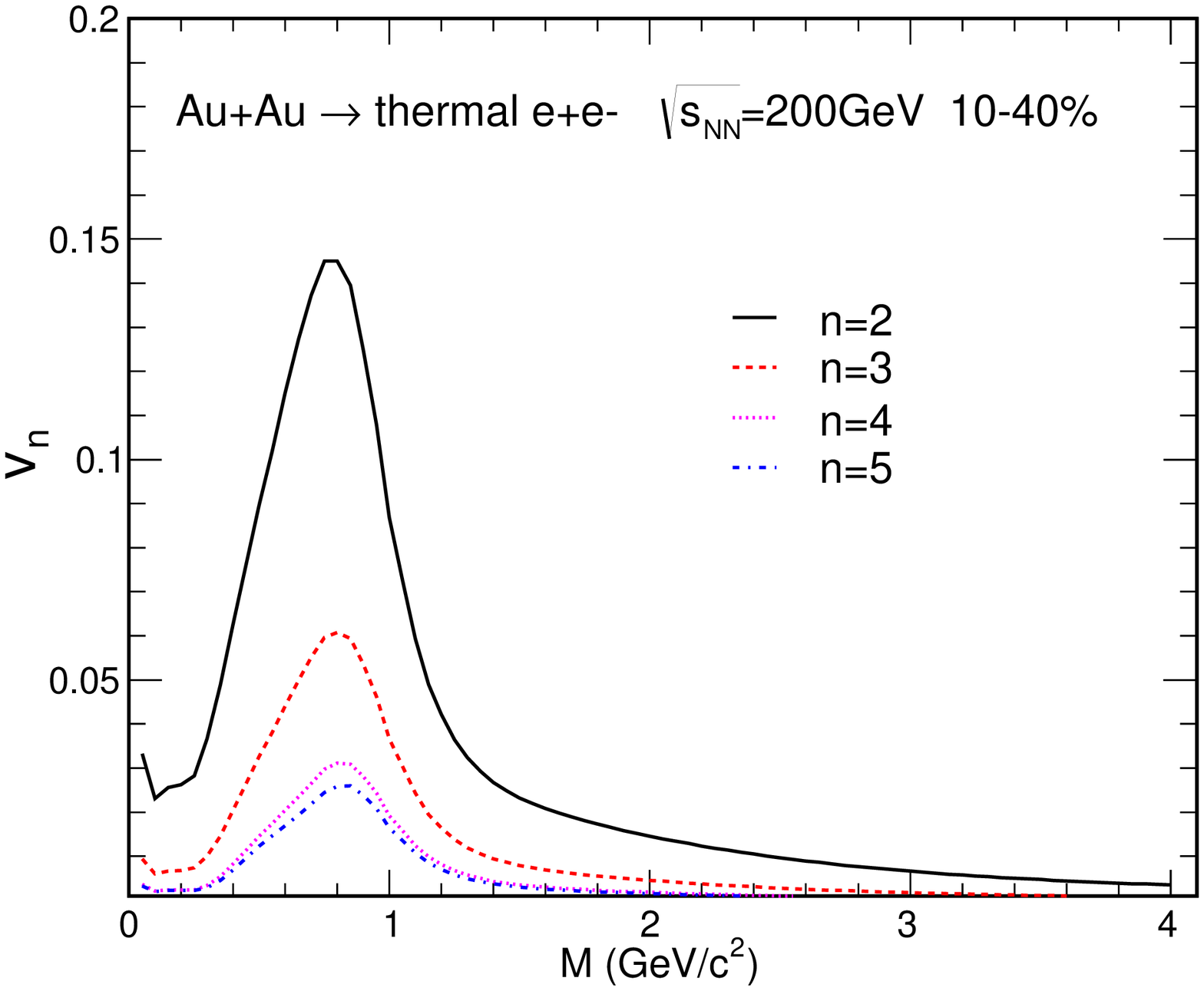}

\protect\caption{\label{fig:vn_10-40} (Color Online) Predicated harmonics coefficients
$v_{n}$ ($n=2,3,4,5$) of thermal dielectrons are shown by various
curves for centrality 10-40\%.}
\end{figure}

In Fig.\ref{fig:vn_10-40}, the higher order harmonics coefficients
$v_{n}$ ($n=2,3,4,5$) of thermal dielectrons from Au-Au collisions
at $\sqrt{s_{NN}}=200$~GeV with centrality of 10-40\% is presented.
The curves have same shape, same trend as direct photons. The magnitude
decreases monotonically with the order $n$. And it is the case for
all the investigated centrality class, 0-10\%, 10-40\% and 40-80\%.
The low $pt$ charged hadrons behaves similarly, as observed by ALICE~\cite{ALICE:2011ab}.

\begin{figure}
\includegraphics[scale=0.4]{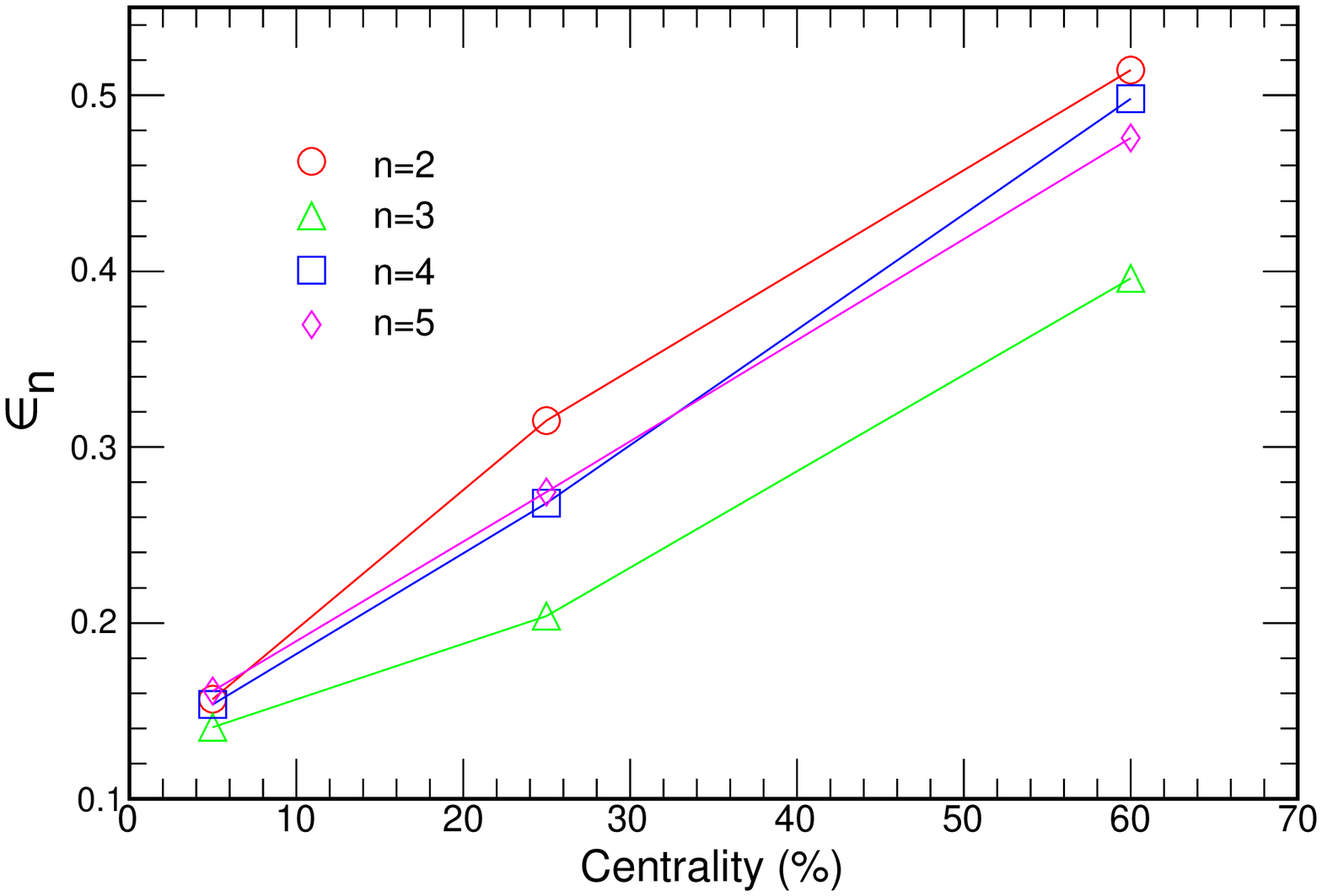}

\protect\caption{\label{fig:ecc_cen1} (Color Online) Eccentricity $e_{n}$ as function
of centrality for $n=2,3,4,5$ with EPOS3.}
\end{figure}

It's often said that the development of anisotropic flow is controlled
by the anisotropies in the pressure gradients which in turn depend
on the shape and structure of the initial density profile. The latter
can be characterized by a set of harmonic eccentricity coefficients
$\epsilon_{n}$ as defined in \cite{Heinz:2013th}. The initial space
eccentricity $\epsilon_{n}$ as a function of centrality for $n=2,3,4,5$
in EPOS3 is shown in Fig.\ref{fig:ecc_cen1}, which looks quite similar
to the initial space eccentricity in IP-Glasma, MC-Glauber and MC-KLN
as shown in \cite{Heinz:2013th}.

What builds up such a large elliptic flow of thermal dileptons, as
shown in Fig.~\ref{v2MB}? The same question is asked in the direct
photons paper based on EPOS3~\cite{Fu2015}. As explained there,
most models predict an elliptic flow too low, whereas EPOS3 can reproduce
the measured elliptic flow. A systematic study~\cite{Fu2015} on
the time evolution of space eccentricity and momentum eccentricity
of the hot dense matter in EPOS3 shows that a correct interplay between
the momentum eccentricity and radial flow is important to account
for the elliptic flow of both hadrons and electromagnetic probes,
dileptons and direct photons. Compared to the surface emission of
hadrons, the overall emission inside the hot dense matter makes direct
photons and dileptons more sensitive to the radial flow. Thus a large
elliptic flow of direct photons and dileptons are obtained at the
same time, with a larger mean radial flow and smaller momentum eccentricity.
It should be said that we use the default EPOS3, with all pamaters
fixed from hadronic observables.

\section{DISCUSSION AND CONCLUSION}

We investigated the anisotropic emission of thermal dileptons from
Au+Au central collision at $\sqrt{s_{NN}}=200$~GeV at the RHIC,
accompanying the study of the direct photons in \cite{Fu2015}, where
a possible solution to the direct photon puzzle is provided.

With a detailed explanation of dilepton emission rate and a relatively
brief explanation of the space-time evolution within EPOS3, we first
calculated the thermal contribution (from QGP and $\rho$ mesons)
to the invariant mass spectra at different centralities, comparing
with the cocktail contribution. The thermal contribution does play
an important role in the low invariant mass region, ie, $M<1GeV/c^{2}$,
which offers a useful window to study the hot and dense matter created
in relativistic heavy ion collisions, especially the early stage of
the hot and dense matter. Our calculated spectra, together with STAR
cocktail, could reasonable well reproduce STAR's measured invariant
mass spectra of di-electron at different centralities.

Then, we investigated elliptic flow and higher order harmonics of
thermal dileptons. The elliptic flow of thermal dileptons is predicted
to be quite large, larger than in other models, and comparable to
the dilepton data for centrality 0-80\%.

Finally, we made predictions of the $v_{n}$ of thermal dileptons
for $2\leq n\leq5$ at three centrality classes, 0-10\%, 10-40\%,
and 40-80\%. The $v_{n}$ of thermal dileptons is found to decrease
with $n$ monotonically for the three centrality class . The same
centrality dependence for $2\leq n\leq5$ is observed with EPOS3,
namely, more central collisions provide smaller $v_{n}$ of thermal
dileptons. Further experimental measurements are highly awaited to
test our solution to the puzzle of direct photon elliptic flow and
improve our understanding of the created hot and dense matter.
\begin{acknowledgments}
S. X. Liu thanks J. Zhao and H. J. Xu for very helpful discussion.
This work was supported by the Natural Science Foundation of China
under Project No.11275081 and by the Program for New Century Excellent
Talents in University (NCET). \end{acknowledgments}

\end{document}